\documentclass[conference]{IEEEtran}

\usepackage{url}
\usepackage[cmex10]{amsmath}
\usepackage{multirow}
\usepackage{epsfig}
\usepackage{mathrsfs}
\usepackage{amssymb}
\usepackage{comment}
\usepackage{bm}
\usepackage{array}
\usepackage{amsthm}
\usepackage{blkarray}
\usepackage{enumerate}
\usepackage{chemarrow}
\usepackage[lined,ruled,linesnumbered]{algorithm2e}
\usepackage{cite}
\usepackage{amsmath,amssymb,amsfonts}
\usepackage{algorithmic}
\usepackage{graphicx}
\usepackage{textcomp}
\usepackage{subcaption}
\usepackage{caption}
\usepackage{makecell}
\usepackage[papersize={8.5in,11in},top=0.75in, bottom=1in, left=0.64in, right=0.64in]{geometry}
\usepackage[labelformat=simple]{subcaption}

\usepackage[dvipsnames]{xcolor}
\definecolor{editing}{rgb}{0, 0, 255}

\newcolumntype{P}[1]{>{\centering\arraybackslash}p{#1}}
\newcolumntype{M}[1]{>{\centering\arraybackslash}m{#1}}

\newtheorem{theorem}{Theorem}

\newtheorem{definition}[theorem]{Definition}

\newcommand{\vars}{\texttt}

\def\BibTeX{{\rm B\kern-.05em{\sc i\kern-.025em b}\kern-.08em
    T\kern-.1667em\lower.7ex\hbox{E}\kern-.125emX}}

\IEEEoverridecommandlockouts\IEEEpubid{\makebox[\columnwidth]{978-1-7281-2700-2/19/\$31.00 2019 $\copyright$ IEEE \hfill}\hspace{\columnsep}\makebox[\columnwidth]{ }}
\begin{document}

\title{MACS: Deep Reinforcement Learning based SDN Controller Synchronization Policy Design\\
\thanks{This research was sponsored by the U.S. Army Research Laboratory and the U.K. Ministry of Defence under Agreement Number W911NF-16-3-0001. The views and conclusions contained in this document are those of the authors and should not be interpreted as representing the official policies, either expressed or implied, of the U.S. Army Research Laboratory, the U.S. Government, the U.K. Ministry of Defence or the U.K. Government. The U.S. and U.K. Governments are authorized to reproduce and distribute reprints for Government purposes notwithstanding any copyright notation hereon.}
}

\author{\IEEEauthorblockN{Ziyao Zhang\IEEEauthorrefmark{1}, Liang 
		Ma\IEEEauthorrefmark{2},  Konstantinos Poularakis\IEEEauthorrefmark{3}, Kin K. Leung\IEEEauthorrefmark{1}, Jeremy Tucker\IEEEauthorrefmark{4}, and Ananthram Swami \IEEEauthorrefmark{5}}
	\IEEEauthorblockA{\IEEEauthorrefmark{1}Imperial College London, London, United Kingdom. Email: \{ziyao.zhang15, kin.leung\}@imperial.ac.uk\\
		\IEEEauthorrefmark{2}IBM T. J. Watson Research Center, Yorktown Heights, NY, United States. Email: maliang@us.ibm.com\\
		\IEEEauthorrefmark{3}Yale University, New Haven, CT, United States. Email: konstantinos.poularakis@yale.edu\\
		\IEEEauthorrefmark{4}U.K. Defence Science and Technology Laboratory, Salisbury, United Kingdom. Email: jtucker@mail.dstl.gov.uk\\
		\IEEEauthorrefmark{5}U.S. Army Research Laboratory, Adelphi, MD, United States. Email: ananthram.swami.civ@mail.mil
}}

\maketitle

\begin{abstract}
In distributed software-defined networks (SDN), multiple physical SDN controllers, each managing a network \emph{domain}, are implemented to balance centralised control,  scalability, and reliability requirements. In such networking paradigms, controllers synchronize with each other, in attempts to maintain a logically centralised network view. Despite the presence of various design proposals for distributed SDN controller architectures, most existing works only aim at eliminating anomalies arising from the inconsistencies in different controllers' network views. However, the performance aspect of controller synchronization designs with respect to given SDN applications are generally missing. To fill this gap, we formulate the controller synchronization problem as a \emph{Markov decision process (MDP)} and apply reinforcement learning techniques combined with deep neural networks (DNNs) to train a \emph{smart, scalable, and fine-grained} controller synchronization policy, called the \emph{Multi-Armed Cooperative Synchronization (MACS)}, whose goal is to maximise the performance enhancements brought by controller synchronizations.  Evaluation results confirm the DNN's exceptional ability in abstracting latent patterns in the distributed SDN environment, rendering significant superiority to MACS-based synchronization policy, which are $56\%$ and $30\%$ performance improvements over ONOS and greedy SDN controller synchronization heuristics.
\end{abstract}


\section{Introduction}
\label{introduction}
Software-Defined Networking (SDN) \cite{kreutz2015software}, a newly-developed  networking architecture, improves network performance due to its programmable network management, easy reconfiguration, and on-demand resource allocation, which has therefore attracted considerable research interests.
One key attribute that differentiates SDN from classic networks is the centralisation of network control, for which all control functionalities are abstracted and implemented in the \emph{SDN controller} sitting in the control plane, for operational decision-making. While the data plane, consisting of \emph{SDN switches}, only passively executes the instructions received from the control plane. Since the logically centralised SDN controller has full knowledge of the network status, it is able to make the global optimal decision. Yet, such centralised control suffers from major scalability and reliability issues. In this regard, \emph{distributed SDN} \cite{bannour2018distributed} is proposed to balance the centralised and distributed controls. 

A distributed SDN network is composed of a set of subnetworks, referred to as \emph{domains}, each managed by a physically independent SDN controller. The physically distributed controllers synchronize with each other to maintain a logically centralised network view, which is referred to as \emph{controller synchronization}. Since complete synchronization among controllers, i.e., all controllers always maintain the same global view, will incur high costs especially in large networks\cite{muqaddas2017inter,qin2018sdn}, most practical distributed SDN networks can only afford partial inter-controller synchronizations and allow temporary inconsistency in controllers' network view, which is known as the \emph{eventual consistency model}\cite{panda2013cap}.  


Under the eventual consistency model, existing works have identified and addressed some serious anomalies arising from controllers' inconsistent network views, such as loopholes\cite{mahajan2013consistent}, blackholes\cite{forster2016consistent}, and other problems caused by policy inconsistencies\cite{reitblatt2012abstractions}. Yet, despite these efforts aimed at eliminating inconsistency-caused anomalies, \emph{we have not seen any notable proposals on fine-grained controller synchronization designs which are tailored for SDN applications with specific performance metrics.} The urgency to fill this gap is especially pronounced when SDN technologies are discussed in a wider range of contexts, where advanced applications are developed on top of SDN-enabled 5G, smart grid, and ISP networks; all these cases require the support of new and finer-grained controller synchronization models\cite{foerster2018survey}. In this regard, we approach the controller synchronization problem with the aim of developing fine-grained controller synchronization policies for enhancing given performance metrics. Complementary to the existing works that make sure the controller synchronization process is error-free; ours is performance-focused, for which we look at which controller(s) should synchronize at certain time  steps so that the given performance metric is maximised. 

To this end, we first define (i) the SDN application of interest whose performance depends on joint communication and computation resources optimisations; and (ii) the corresponding performance metric. Then, we formulate the problem of developing the controller synchronization policy that maximises the defined performance metric as a \emph{Markov decision process (MDP)}, which can be solved by employing reinforcement learning (RL) techniques. RL-based approaches are especially appealing for developing the controller synchronization policy under distributed SDN for the following reasons. (i) The abundance of network data made available by SDN switches through the OpenFlow protocol \cite{amaral2016machine} builds up a pool of past experiences which are the ideal ``trial-and-error" inputs for RL algorithms. (ii) Different SDN domains can be highly heterogeneous; as such,  SDN networks are complex systems. Therefore, accurately modelling such systems becomes extremely difficult and mathematically intractable. In light of this, the model-free RL-based approaches are especially attractive, as they come without any constraints on network's structure or its dynamicity, thus adaptable for handling real-world SDN networks. Inspired by its recent successes, we propose a deep reinforcement learning (DRL)-based controller synchronization framework, which employs deep neural networks (DNN) to generalise synchronization policy estimations, called Multi-Armed Cooperative Synchronization (MACS) to assist controllers in learning synchronization policies based on past experiences.  
Since the synchronization decision at one time instance could have lasting effects, the goal of MACS is to maximise the long-term performance enhancement brought by controller synchronizations with respect to (w.r.t.) the given  metric.  \looseness = -1

Extensive evaluations show that MACS, operating on limited controller synchronization budgets, is nearly as good as the full synchronization scenario where all distributed controllers are always synchronized. Overall, MACS outperforms the default anti-entropy synchronization algorithm implemented in ONOS controllers and a greedy heuristic by up to $56\%$ and $30\%$, respectively. To the best of our knowledge, MACS is the first DRL-based SDN controller synchronization scheduler which yields fine-grained synchronization policies aiming at optimising both communication and computation resources. 

The rest of the paper is organised as follows. Section~\ref{sec:sys_description} describes the distributed SDN system. Section~\ref{sec:problem_formulation} formulates the SDN controller synchronization problem as an MDP. Section~\ref{sec:dq_scheduler} presents design details of MACS. Section~\ref{sec:evaluation} presents evaluation results and analysis. Section~\ref{sec:related_work} discusses related work. Finally, Section~\ref{sec:conclusion} concludes the paper. 

\section{System Description}
\label{sec:sys_description}
This section briefly describes the distributed SDN environment and the SDN application of interest.


\subsection{SDN Controller}
\label{sdn_controller}
Under distributed SDN architectures, all control functionalities are abstracted and implemented in distributed SDN controllers, which are responsible for their designated domains (referred to as \emph{domain controllers}), for operational decision-makings. Domain controllers have full and up-to-date view of their designated domains; they synchronize with each other (Section~\ref{controller_sync}) to a maintain logically centralised network view for better network performance (see the example in Section~\ref{controller_sync} for the role of controller synchronization). Routine and frequent control tasks are handled by domain controllers in individual domains, based on their network views. 
In addition, we assume that there exists a \emph{central controller}, whose only responsibility is to develop synchronization policies and coordinate controller synchronizations, based on information supplied by domain controllers. Such a central controller could be one of the existing distributed controllers, or an independent control unit. All controllers therefore forms the \emph{SDN control plane}.
\looseness = -1

\subsection{SDN Domain}
\label{sdn_domain}
In this paper, an SDN domain refers to the collection of network elements managed by a domain controller. These elements may include SDN switches, servers, and users that are connected to the SDN switches.
One important advantage brought by the SDN paradigm is the ability to virtualise network services and install them on general purpose \emph{servers} inside SDN domains. This is in clear contrast to the traditional networking paradigm where network services are provided by dedicated equipments with designated functionalities running on specific protocols and proprietary configuration tools.
Each domain contains one or multiple gateway routers (or switches, in this paper, they are all referred to as \emph{gateway routers}) connecting to other SDN domains. 

\subsection{Controller Synchronization Application}
\label{application}
To materialise potential performance gains controller synchronization can bring under distributed SDN, we focus on an application where find-grained information about communication and computation resources are essential for its operation. Therefore, we choose \emph{service path construction} as the application of interest. In the Network-as-a-Service (NaaS) SDN environment, QoS-aware service path construction is a crucial problem in the context where network services are virtualised in servers \cite{duan2012survey}. Specifically, we investigate the problem where several network services, e.g., wireless access admission, firewall, etc., are installed on servers across all SDN domains. Requests for services are submitted by users to domain controllers, who construct service paths for requests submitted, based on their network views. Note that the process of finding a service path is an anycast problem\cite{abley2006operation}, as a service can have multiple installations in different domains. In order to calculate the best service path w.r.t. the given performance metric, domain controllers rely on up-to-date information about other domains (e.g., traffic levels, network delay, available computation and services, etc.) gained through controller synchronization. 


\subsection{Network Model}
\label{network_model}
We formulate the distributed SDN network as a directed graph, where $m$ vertices are connected via directed links. Let the graph representing the SDN network be denoted by $\mathcal{G}=(V,E)$ ($V$/$E$: set of vertices/edges in $\mathcal{G}$, $|V|=m$), which is referred to as the \emph{domain-wise topology}. The existence of two edges in $e_{1,2} , e_{2,1} \in E$ connecting two vertices $v_1,v_2\in V$ in the domain-wise topology implies that the two network domains corresponding to $v_1$ and $v_2$ are connected. Then, we further associate weights for the directed inter-domain links, which represent \emph{gateway delays} (see Section~\ref{performance_metric}). We use $\mathcal{L}_{i}$ to denote the set of outgoing edge weights from the vertex of domain $\mathcal{A}_{i}$ in the network. For instance, let $\mathcal{A}_{1}$ and $\mathcal{A}_{2}$ denote the network domains corresponding to vertices $v_1$ and $v_2$ in the graph; then the weight of edge $e_{1,2}$, denoted by $l_{1,2} \in \mathcal{L}_{1}$, represents the estimation of latency a packet going thorough domain $\mathcal{A}_{1}$ and entering domain $\mathcal{A}_{2}$ should expect to experience in $\mathcal{A}_{1}$ (i.e., the \emph{gateway delay}); similarly, $l_{2,1} \in \mathcal{L}_{2}$ is the latency a packet going thorough domain $\mathcal{A}_{2}$ and entering domain $\mathcal{A}_{1}$ would experience in $\mathcal{A}_{2}$. 

Let $\mathcal{C}$ be the set of all installed services on servers located across different domains in the SDN network. Let $c^{(i)}_{j}$ denote service $i$ installed in domain $\mathcal{A}_j$. Note that we do not differentiate two identical services installed in the same domain. Then, $\mathcal{D}_{j}$ is the set of server delays of all service installations in domain $\mathcal{A}_{j}$, and  $d_{j}^{(i)} \in \mathcal{D}_{j}$ denotes the waiting time before a request for service $i$ starts being processed in domain $j$ (i.e., the \emph{server delay}, see Section~\ref{performance_metric}).

\emph{\textbf{Discussion:}} Note that due to network dynamicity, e.g., traffic levels, user demand for service patterns, etc., the values of $l_{i,j}$ and $d_{j}^{(i)}$ are time-varying. With the SDN settings described in this section, domain controllers always have the up-to-date status views of network elements residing in their domains; in other words, the controller of domain $\mathcal{A}_{i}$ always knows the newest $\mathcal{L}_{i}$ and $\mathcal{D}_{i}$. The controller of domain $\mathcal{A}_{i}$ relies on synchronizations with other domain controllers (see Section~\ref{controller_sync}) to learn up-to-date $\mathcal{L}_{j} (j\neq i)$ and $\mathcal{D}_{j}(j\neq i)$.

\section{Problem Formulation}
\label{sec:problem_formulation}
In this section, 
we first define the performance metric, controller synchronizations and the synchronization budget. Then, we discuss the service path construction mechanism employed, which is followed by the MDP formulation of our controller synchronization problem. 

\subsection{Performance Metric}
\label{performance_metric}
For the user who submitted a service request, the sooner the request gets served, the better. Therefore, the gap in time between the submission of a service request and the server starting processing the request is a natural performance measurement of qualities of constructed service paths, which we call the \emph{request latency}. 

Request latencies consist of two parts: transit latency and the waiting time at the server. Many factors, such as the link congestion levels, the number of hops, may contribute to the overall transit latency for the inter-domain routing of a service request from the user domain to the domain of the chosen server. For easier modelling, we use the delays incurred at egress gateway routers of SDN domains (referred to as \emph{gateway delays}), which are usually the bottlenecks \cite{das2007mitigating}, to abstract the transit latency incurred traversing through SDN domains. If a service request is submitted to a server located in the same domain as the user, transit latency incurred is assumed to be negligible.
On the other hand, delays incurred at the server are the waiting times in the server before available computation resources can be assigned for processing the submitted requests, which we refer to as the $\emph{server delay}$.  Based on the network model described in Section~\ref{network_model}, $l_{i,j}$ and $d_{j}^{(i)}$ correspond to gateway and server delays, respectively.

Therefore, we define the performance metric w.r.t. a constructed service path as the accumulated gateway delays en route the service path and server delay at the chosen server, which is referred to as \emph{request latency} in the sequence. 

\subsection{Synchronization Among SDN Controllers}
\label{controller_sync}
\begin{figure}
	\smallskip
	\centering
	\includegraphics[height=0.25\linewidth,width=1.1in]{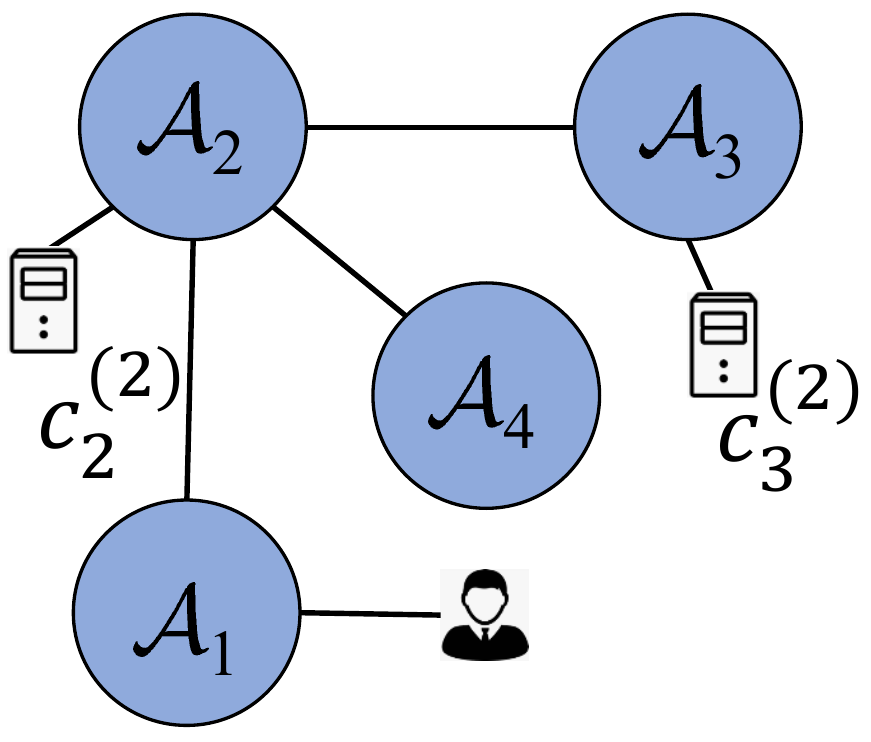}
	\caption{A service path construction example.}
	\label{fig:service_path_example}
	\vspace{-2em}
\end{figure}


W.r.t. the application of interest introduced in Section~\ref{application}, controller synchronization levels directly affect the quality of constructed service paths. We use an example to demonstrate this.
In Fig.\ref{fig:service_path_example}, suppose the user located in domain $\mathcal{A}_{1}$ submits a service request for service $2$. The domain controller for $\mathcal{A}_{1}$ is aware that the request can be forwarded to the server hosting $c^{(2)}_{2}$ in domain $\mathcal{A}_{2}$ or the server hosting $c^{(2)}_{3}$ in domain $\mathcal{A}_{3}$. The domain controller uses its view of the network conditions in domains $\mathcal{A}_{2}$ and $\mathcal{A}_{3}$ to decide the service path which it thinks would minimise the request latency before the request can be served. However, the domain controller's view could be stale. Specifically, domain $\mathcal{A}_{1}$ controller's view of the gateway delay between $\mathcal{A}_{1}-\mathcal{A}_{2}$, $\mathcal{A}_{2}-\mathcal{A}_{3}$ are $2$ and $1$, respectively. $\mathcal{A}_{1}$'s view of the gateway delay between $\mathcal{A}_{1}-\mathcal{A}_{2}$ is always accurate and up-to-date, since the gateway router is within domain $\mathcal{A}_{1}$; while the up-to-date gateway delay between $\mathcal{A}_{2}-\mathcal{A}_{3}$ is actually $3$. On the other hand, $\mathcal{A}_{1}$'s view and the up-to-date server delays of the servers hosting $c^{(2)}_{2}$ and $c^{(2)}_{3}$ are $4$ and $2$, $1$ and $3$, respectively. Therefore, the domain controller of $\mathcal{A}_{1}$ would forward the service request to domain $\mathcal{A}_{3}$ rather than $\mathcal{A}_{2}$ since the request latencies are estimated to be $2+1+2 = 5$ in the former and $2+4 = 6$ in the latter based on its view of the network. However, if $\mathcal{A}_{1}$'s domain controller holds up-to-date view of the network condition in domains $\mathcal{A}_{2}$ and $\mathcal{A}_{3}$, it would send the request to domain $\mathcal{A}_{2}$ instead. 

This example highlights the important role of controller synchronization that distributes up-to-date network information among domain controllers. Here, we formally define controller synchronization w.r.t. our service path construction problem in this subsection. First, we define the unit that quantises the synchronizable domain information. 
\vspace{-.5em}
\begin{definition}
	\label{def:BIS}
	The gateway delay between a pair of domains, or the server delay of a service is referred to as a Basic Information of Synchronization (BIS). 
	\vspace{-.5em}
\end{definition}
A BIS corresponds to the most fundamental piece of information that can be synchronized to domain controllers. Note that in existing distributed controller implementations such as ONOS, when two controllers synchronize, they exchange their entire state information. In this paper, we propose a more fine-grained synchronization policy where only selected state information is exchanged. According to the above definition, an SDN network $\mathcal{G}$, with the set of installed service $\mathcal{C}$, has $N = |E| + |\mathcal{C}|$ number of BISes in total (see Section~\ref{network_model}). With the concept of BIS, we then formally define controller synchronization in the following definition. 
\looseness = -1
\vspace{-.5em}
\begin{definition}
	\label{def:synch}
    Controller synchronization is the process of domain controllers broadcasting/receiving up-to-date BISes originated in their domains/received from other domain controllers.
    \vspace{-1.3em}
\end{definition}
Definition~\ref{def:synch} implies that controllers synchronize with each other in the way that selected up-to-date BIS(es) are broadcasted to all domain controllers via control plane messages. Then, all domain controllers update their network views by incorporating the received up-to-date BIS(es) from other controllers and BIS changes reported in their own domains.

As discussed briefly in Section~\ref{introduction}, frequent dissemination of inter-controller synchronization messages introduce prohibitively large overheads\cite{muqaddas2017inter,qin2018sdn}. Moreover, excessive status updates can potentially lead to network performance degradation caused by instability \cite{labovitz1998internet}. Indeed, we show in Section~\ref{evaluation_results} that excessive synchronizations cause performance deterioration for MACS. Therefore, the number of synchronization messages that can be exchanged at a time is limited, for which we introduce the \emph{synchronization budget}.
\vspace{-.5em}
\begin{definition}
	\label{def:sync_budget}
	The synchronization budget at a time is the maximum number of BISes that can be broadcasted simultaneously.
	\vspace{-1.5em}
\end{definition}

\emph{\textbf{Discussion:}} Note that in addition to the synchronizable BIS, 
all domain controllers always have knowledge of the correct domain-wise topology (without edge weights) and the available services in every other domains.

\subsection{Service Path Construction using BIS Information}
\label{path_construction}

Due to the flexibility and programmability of the SDN, there are potentially many ways in which domain controllers calculate service paths. In this section, we describe a simple service path construction mechanism that uses BISes and aims at minimising the overall request latency. Note that in this paper, it is \emph{not} our intention to design any new such mechanisms; we use this simple and representative mechanism for the sake of problem formulation. Specifically, the service path construction mechanism consists of the following steps.

\emph{\textbf{Step 1}}: The domain controller that receives a service request (\emph{source controller} in the sequence) calculates the minimum accumulated gateway delay(s) of service path(s) leading to all possible server(s) that host the requested service (recall that the process of finding a server is an anycast problem). 

\emph{\textbf{Step 2}}: The domain controller calculates the request latency (latencies) for the requested service for all potential service path(s) in \emph{\textbf{Step 1}} by combining their accumulated gateway delays and the corresponding server delays. The source controller chooses the service path that incurs the lowest request latency based on its calculation results. 
 
\emph{\textbf{Step 3}}: The forwarding rule of the service request is installed on involved SDN switches in forms of flow table entries. \looseness = -1

Applying the above service path construction mechanism, only the source controller decides the domain-wise service path based on its view of the network status. Other domain controllers are deliberately left out in calculating service paths to avoid forwarding anomalies, e.g., routing loops and black holes, which could arise if different domain controllers with heterogeneous network views attempt to independently calculate service path for a packet transiting through their domains. \looseness = -1

\subsection{The Objective of Controller Synchronization Policy}
\label{sec:objective}


Two questions motivate our definition of the objective of controller synchronization. First, how does the central controller develop the synchronization policy that maximises the performance metric, given the limited synchronization budget? Second, since a synchronization decision has lasting effects, how does the synchronization policy maximise the performance enhancement of controller synchronization over time? Here, a \emph{synchronization policy} refers to a series of synchronization decisions (i.e., which up-to-date BIS(es) to broadcast, subject to the available synchronization budget) over a period of time. With this in mind, we state the objective below. 

\emph{\textbf{Objective}}: In dynamic networks where gateway and server delays are time-varying, given the controller synchronization budget, 
how do domain controllers synchronize with each other by broadcasting up-to-date BISes, to maximise the performance improvements brought by controller synchronizations (i.e., reductions in average request latency due to the availability of accurate BISes via synchronizations) over a period of time?  

\subsection{Markov Decision Process (MDP) Formulation}
\label{sec:MDP_formulation}
MDP \cite{white2001markov} offers a mathematical framework for modelling serial decision-making problems. Specifically, our controller synchronization policy can be modelled as an MDP with the 3-tuple $(\mathcal{S},\mathcal{A},R)$ as follows.


\begin{itemize}
	\item  $\mathcal{S}$ is the finite state space. In particular, a state corresponds to the collection of the respective counts of time slots elapsed since the last broadcasts of up-to-date values of each BIS. As such, a state of the MDP represents the \emph{staleness} of status information of network components. The size of a state is $N$, which is the total number of BISes. \looseness = -1
	\item $\mathcal{A}$ is the finite action space. An action w.r.t. a state is defined as whether each up-to-date BIS should be broadcasted (indicated by $1$) or not (indicated by $0$), i.e., $\mathcal{A} \in \{1,0\}^{N}$. As such, the size of an action is also $N$. 
	\item $R$ represents the immediate reward associated with state-action pairs, denoted by $R(\mathbf{s},\mathbf{a})$, where $\mathbf{s} \in \mathcal{S}$ and $\mathbf{a} \in \mathcal{A}$ are the state and action vectors. $R(\mathbf{s},\mathbf{a})$ is calculated as the reductions in average request latency of all service requests in the network after taking action $\mathbf{a}$ in state $\mathbf{s}$.
\end{itemize}

With this MDP formulation,  $\mathcal{S}$ and $R$ are collected by domain controllers from data planes through SDN's northbound interface\cite{kreutz2015software} and are supplied to the central controller.

\begin{figure}
	\smallskip
	\centering
	\includegraphics[height=0.5\linewidth,width=2.2in]{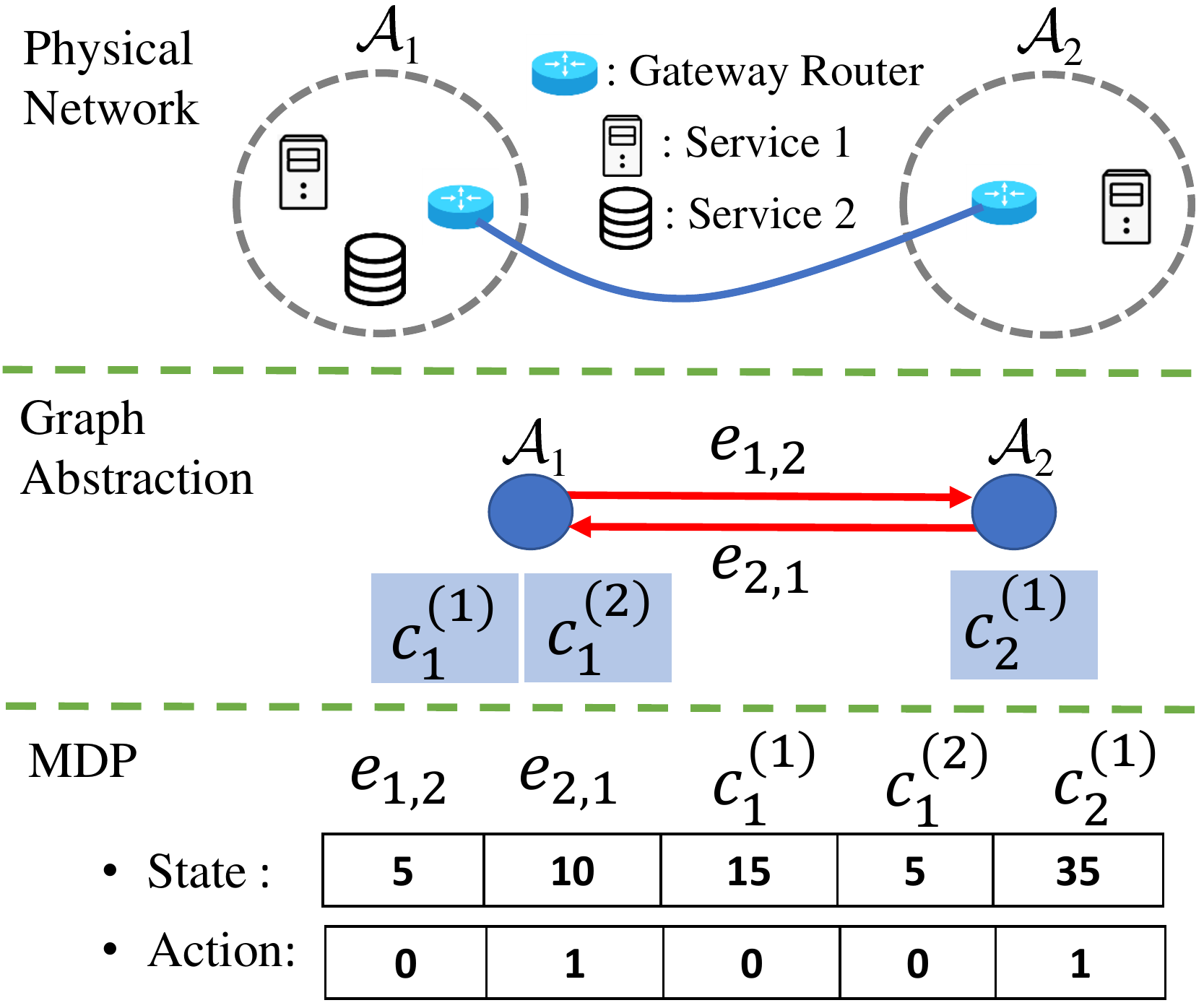}
	\caption{An SDN network with $2$ domains $\mathcal{A}_{1}$ and $\mathcal{A}_{2}$, where service $1$ is available in both domains and service $2$ is only available in $\mathcal{A}_{1}$. Two domains are connected by a pair of gateway routers. There are in total $5$ BISes in the network.} 
	\label{fig:state_action_example}
	\vspace{-1.5em}
\end{figure}

Fig.~\ref{fig:state_action_example} demonstrates the  $(\mathcal{S},\mathcal{A})$ pair in the formulated MDP in a simple example. 
The first entry in the state vector indicates that the last synchronization broadcast of the up-to-date value of gateway delay from  $\mathcal{A}_{1}$ to $\mathcal{A}_{2}$ took place 5 time slots ago. An action consisting of binary entries indicates that the up-to-date value of gateway delay from $\mathcal{A}_{2}$ to $\mathcal{A}_{1}$ and the server delay of service $1$ installed in domain $\mathcal{A}_{2}$ are to be broadcasted to all domain controllers at the current time slot.

The \emph{optimal action} at each state is defined as the action that yields the maximum long-term reward, which reflects the fact that the synchronization decision at a time slot has lasting effects. In particular, the long-term reward is defined as the discounted sum of the expected immediate reward of all future state-action pairs from the current state. The reward for the state-action pair $\Delta t$ steps ahead of the current state is discounted by $\gamma^{\Delta t}$, where $\gamma$ is called the \emph{discount factor} and $0<\gamma<1$. Here, $\gamma$ trades off the importance between the current and the future reward. Therefore, starting from an initial state $s_0$, the problem is formulated as finding a policy $\pi$ (i.e., the selection of a sequence of actions ${\{\mathbf{a}_{t}\}}_{t=0}^{T}$) such that the long-term accumulated reward expressed in the Bellman equation below is maximised
\vspace{-.5em}
\begin{equation}\label{eq:long_term_reward}
V^{\pi}(\mathbf{s}_0) = \mathbb{E}_{\pi}\bigg[\sum_{t=1}^{T}\gamma^t R(\mathbf{s}_{t},\mathbf{a}_{t})|\mathbf{s}_0\bigg],
\vspace{-.5em}
\end{equation}
where $\mathbf{s}_t$ and $\mathbf{a}_t$ are the state-action pair at time $t$, and $T$ is the total time horizon of the problem. 

\emph{\textbf{Time Scale of the MDP:}} The basic unit of time in the defined MDP include a series of events, which are jointly referred to as a \emph{synchronization time slot} (\emph{time slot} for short) in the sequence. Specifically, synchronization broadcastings take place at the start of a time slot. After all domain controllers finish broadcasting and receiving up-to-date BIS, they recalculate service paths for requests originated from their respective domains, based on the updated network views. Then, actual service latencies en route for new service paths are recorded for the calculation of $R$. For modelling tractability, we assume that newly broadcasted BIS values remain accurate until actual service latencies are recorded. This is true in practical SDN networks where periodical synchronization means that the collection of latest network information takes place at certain time. The collected information is considered to be ``up-to-date" for a short while, as ``real time" is a relative concept. It should be stressed that the concept of ``times slot'' in the formulated MDP is very different to the performance metric, i.e., service latency. The former is defined as a full cycle of controller synchronization after which the MDP shifts from one state to another state. In contrast, the latter, which is coupled with time, is the actual length of a request waiting time. Since domain controllers need to record the actual service latency before the end of an MDP time slot, the length of a time slot is determined by the latency incurred en route the most delayed service path constructed at the beginning of the time slot.
\looseness = -1

\emph{\textbf{Discussion:}} Note that our MDP formulation is not only specific to the service path construction application defined in 
Section~\ref{application}. In essence,  the definition of the state space here represents a staleness measure of status information about various network components in the distributed SDN. In our problem, these components are gateway routers and servers. For other controller synchronization problems, as long as networking elements can be itemised to fit in the state definition, such problems can be modelled by the formulated MDP.\looseness = -1

\section{The MACS}
\label{sec:dq_scheduler}
To solve the formulated MDP, we use RL techniques to find the sequence of actions that maximises the Bellman equation in (\ref{eq:long_term_reward}). For RL, imagine an agent who jumps from state to state in the formulated MDP by taking some actions associated with certain rewards. The agent's goal is to discover a sequence of state-action pairs, i.e., a \emph{policy}, that maximises the accumulated time-discounted rewards. By interacting with the environment modelled by the MDP, the agent's experiences build up through ``trial-and-error" where good decisions are positively enforced by positive rewards, and bad ones the opposite. 

During training, the most important aspect is how the agent generalises and memorises what it has learned. Traditionally, the agent's estimations of future reward following state-action pairs are kept in tabular fashion. However, this approach soon becomes impractical in most RL tasks because of large state-action space sizes. Indeed, the state-action space is enormous in our controller synchronization problem. Consider a scenario with $N$ number of BIS and a time horizon of $T$ time slots, then there are as many as $T^{N}$ states and $2^{N}$ actions associated with each state. In light of this, function approximators \cite{bertsekas1995neuro} have been proposed to approximate the $Q$-function, which represents the agent past experiences as it estimates the potential value of a given state-action pair (see Section~\ref{q_learning}). 
Among these approximators, DNN \cite{hagan1996neural} stands out
due to its exceptional ability in capturing latent and complicated relationships from input data. Therefore, we also employ DNN in MACS to help the agent make sense of and generalise past experiences. 

In this section, we first discuss design challenges and the MACS architecture in Section~\ref{chanlleges} and Section~\ref{sec:learning_architecture}, respectively. Then, mathematical details of how policies are estimated and the weights update process for the DNN in MACS are discussed in Section~\ref{q_learning}. Finally, we present the training algorithm for MACS in Section~\ref{training_algorithm}.

\subsection{Challenges}
\label{chanlleges}
From our experiences leveraging several RL techniques to solve the formulated MDP, we identify some non-trivial challenges arising mainly from the two following aspects. First, the problem is a discrete control problem. This prevents the application of a number of well-established and relatively mature actor-critic approaches \cite{konda2000actor} based on the policy gradient theorem \cite{sutton2000policy}. For example, DeepMind's recent work on the deep deterministic policy gradient (DDPG) agent \cite{lillicrap2015continuous} is the state-of-the-art for continuous control problems. 
Second, our formulated MDP has a high-dimensional state-action space. In particular, there are up to $2^{N}$ number of possible actions for any state. Thus, the size of the action space increases exponentially with the number of BISes in the network. It has been shown that such large action space is very difficult to explore and generalise from \cite{lillicrap2015continuous}. Indeed, classic RL techniques \cite{watkins1992q} and their variations \cite{hessel2018rainbow}, which work well in scenarios with relatively small discrete action spaces, are not suitable solutions for our problem. 
\looseness = -1

\subsection{The MACS Architecture}
\label{sec:learning_architecture}
Considering the challenges discussed in the previous section, we need a DRL solution that is designed for discrete problems and can perform well in the presence of enormous state-action spaces. To this end, we build our learning architecture based on proposals in \cite{wang2015dueling,tavakoli2018action}, which have design features suitable for the nature of our formulated MDP. 

DeepMind's dueling network architecture \cite{wang2015dueling} explicitly separates the training for estimations of state-values and the advantages for individual actions, i.e., these values can be obtained separately. This is in contrast to most existing DRL architectures where the output of the DNN is conventionally a single value, i.e., the estimated $Q$-value for the input state-action pair. The dueling network architecture is particularly helpful in situations where there are many similar-valued actions. 
\looseness = -1

The action branching architecture (ABA) \cite{tavakoli2018action} takes the dueling network a step further by categorising all actions as belonging to an \emph{action dimension}. Moreover, a separate action advantage estimator is assigned to each action dimension (referred to as an \emph{action arm} in the neural network) to estimate the advantage of all actions belonging to the action dimension (referred to as \emph{sub-actions}). Under such arrangements, the $Q$-value estimation for each action is obtained by combining (1) the state value estimation, which is shared by all possible actions given the state; and (2) the sub-action advantage estimated by the assigned action arm. A key characteristic of the action branching architecture is that a degree of freedom is given to each action dimension by dedicating a separate arm in the network for advantage estimations of sub-actions belonging to that action dimension. This design greatly improves learning efficiency and reduces complexity which arises from the combinatorial increase of the total number of action dimensions. Therefore, the design principles in the ABA are suitable for approximating the $Q$-function in our formulated MDP, as they address the challenges discussed in Section~\ref{chanlleges}. In the following, we use the example in Fig.~\ref{fig:state_action_example} to demonstrate how these design principles work in MACS. 

\begin{figure}
	\smallskip
	\centering
	\includegraphics[height=0.4\linewidth,width=2.7in]{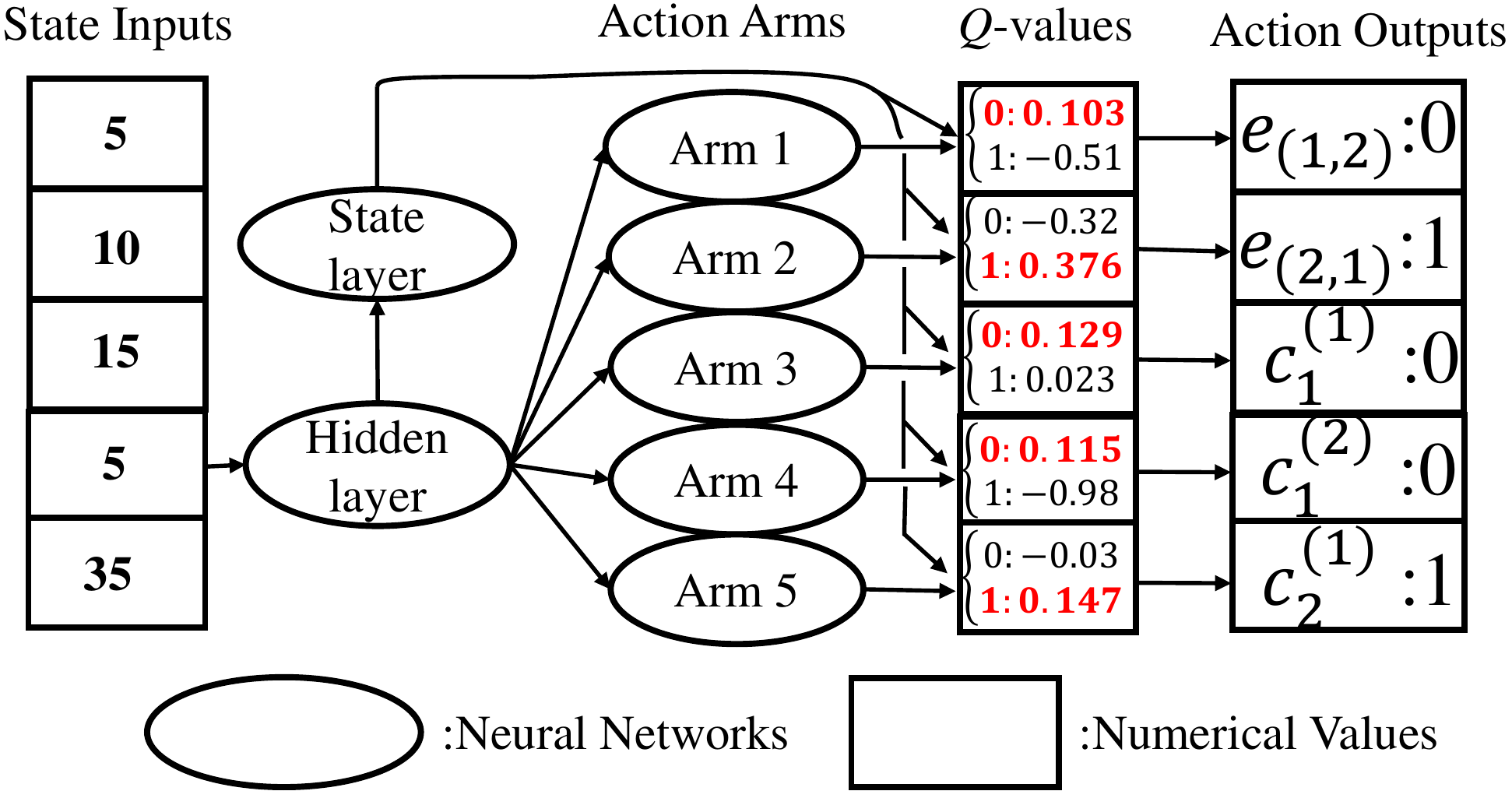}
	\caption{The MACS network based on the example in Fig.~\ref{fig:state_action_example}. State inputs are first fed to the hidden layer, which is shared by the state layer and all action arms. The state layer is responsible for estimating the shared state value, whereas all action arms are responsible for estimating advantages of their $2$ sub-actions. }
	\label{fig:DRL_example}
	\vspace{-1.8em}
\end{figure}

The diagram of the DNN constructed for MACS shown in Fig.~\ref{fig:DRL_example} is based on the example scenario in Fig.~\ref{fig:state_action_example}, where there are 5 BISes. In Fig.~\ref{fig:DRL_example}, 
there are $5$ action arms, each corresponds to a BIS. Moreover, there are two sub-actions in every action dimension, i.e., $0$: not to broadcast the corresponding up-to-date BIS value; and $1$: to broadcast the corresponding up-to-date BIS value. An action arm outputs the estimation of action advantages of the two sub-actions under that action dimension. Furthermore, there is a separate state layer which outputs the estimation of the state value given the state inputs. Both the state layer and all semi-independent action arms are preceded by the input layer, which is designed for coordination. Based on the estimated $Q$-values, which are obtained by combining the estimated state value (output of the state layer) and action advantages (outputs of action arms), the sub-action with the highest $Q$-value is selected for each action arm. Finally, the outputs of all action arms concatenate and form the chosen action w.r.t. the state input.

In the following section, we give further details on how state values, action advantages, and $Q$-values are calculated and their relationships in MACS. 

\subsection{Design Details of MACS}
\label{q_learning}
\emph{$Q$-function}, which originates from the classic \emph{$Q$-learning} algorithm \cite{watkins1992q}, is commonly used to estimate the quality (i.e. potential value) of a state-action pair, i.e., 
$
\mathcal{S} \times \mathcal{A} \rightarrow \mathbb{R}.
$
\smallskip
In particular, the $Q$-function for a state-action pair following a policy $\pi$ is defined as
\vspace{-.5em}
\begin{equation}
\label{eq:Q_function}
Q^{\pi}(\mathbf{s},\mathbf{a}) = \mathbb{E}_{\mathbf{s}^{\prime}}\big[R(\mathbf{s},\mathbf{a})+\gamma\mathbb{E}_{\mathbf{a}^{\prime} \sim\pi(\mathbf{s}^{\prime})} [Q^{\pi}(\mathbf{s}^{\prime},\mathbf{a}^{\prime})]\big],
\vspace{-.3em}
\end{equation}
where $(\mathbf{s}^{\prime},\mathbf{a}^{\prime})$ is the state-action pair at the next step. $Q^{\pi}(\mathbf{s},\mathbf{a})$ estimates the long-term value of a particular state-action pair following policy $\pi$; whereas the \emph{state value}, denoted by $V^{\pi}(\mathbf{s})$, estimates the expected long-term value of the state $\mathbf{s}$
\vspace{-.3em}
\begin{equation}
\label{eq:value_function}
V^{\pi}(\mathbf{s}) = \mathbb{E}_{\mathbf{a}\sim\pi(\mathbf{s})}[Q^{\pi}(\mathbf{s},\mathbf{a})].
\vspace{-.3em}
\end{equation}   

Furthermore, to better distinguish the relative qualities of all possible actions under a given state $\mathbf{s}$, we define the \emph{advantage} of an action $\mathbf{a}$ following policy $\pi$, denoted by $A^{\pi}(\mathbf{s},\mathbf{a})$, as
\vspace{-.5em}
\begin{equation}
\label{eq:action_advantage}
A^{\pi}(\mathbf{s},\mathbf{a}) = Q^{\pi}(\mathbf{s},\mathbf{a})-V^{\pi}(\mathbf{s}).
\vspace{-.5em}
\end{equation}   
W.r.t. the action branching architecture we employ, let 
$a_{i}\in\mathbf{\Omega}_{i}, i\in\{1,\dots,N\}$ denote a sub-action belonging to action arm $i$, where $\mathbf{\Omega}_{i}$ is the set of all sub-actions of action arm $i$. Then, the $Q$-value and the action advantage of $a_{i}$ are denoted by $Q^{\pi}_{i}(\mathbf{s},a_{i})$ and $A^{\pi}_{i}(\mathbf{s},a_{i})$, respectively. A straightforward way to combine the state value and action advantages for $Q$-values is to follow (\ref{eq:action_advantage}), i.e., $Q^{\pi}_{i}(\mathbf{s},a_{i}) = V^{\pi}(\mathbf{s}) + A^{\pi}_{i}(\mathbf{s},a_{i})$. However, as suggested by authors in \cite{tavakoli2018action}, normalising the action advantage by the mean of action advantages before combining it with the shared state value yields better performance. Thus, the following aggregation method is used instead
\vspace{-.3em}
\begin{equation}
\label{eq:aggregation_method}
Q^{\pi}_{i}(\mathbf{s},a_{i}) = V^{\pi}(\mathbf{s}) + \big(A^{\pi}_{i}(\mathbf{s},a_{i})-\frac{1}{|\mathbf{\Omega}_{i}|}\sum_{a_{i}\in \mathbf{\Omega}_{i}}A^{\pi}_{i}(\mathbf{s},a_{i})\big).
\vspace{-.3em}
\end{equation}  

For action arm $i$, since we use DNN as the function approximator of its sub-action's $Q$-function $Q^{\pi}_{i}(\mathbf{s},a_{i})$, it is parametrised by the set of adjustable parameters $\boldsymbol{\theta}_{i}$ which are weights of the DNN. Then, the parametrised $Q$-function is denoted by $Q^{\pi}_{i}(\mathbf{s},a_{i};\boldsymbol{\theta}_{i})$.  The \emph{value iteration update}  \cite{watkins1992q} of the $Q$-function uses the estimation of future rewards at the next state to update current $Q$-function, with the reasoning that estimations at the next state are more accurate, hence increasingly-more-accurate $Q_{i}(\mathbf{s},a_{i};\boldsymbol{\theta}_{i})$ is eventually able to converge to the optimal policy $\pi^{*}$ after enough rounds of iterations. During weight update, $\boldsymbol{\theta}_{i}$ is adjusted to reduce the gap between current prediction (i.e., current $Q_{i}(\mathbf{s},a_{i};\boldsymbol{\theta}_{i})$) and the next state estimate. Specifically, we define the target for arm $i$ as
\looseness = -1
\vspace{-.3em}
\begin{equation}\label{eq:target}
y_{i} = R(\mathbf{s},\mathbf{a})+\gamma\max_{a^{\prime}_{i} \in \mathbf{\Omega}_{i}}Q_{i}(\mathbf{s}^{\prime},a^{\prime}_{i};\boldsymbol{\theta}_{i}).
\vspace{-.5em}
\end{equation}

Then, the following loss function using the mean-squared error measurement is defined for adjusting $\boldsymbol{\theta}_{i}$
\vspace{-.5em}
\begin{equation}\label{eq:Q_update}
\begin{split}
L(\boldsymbol{\theta_{i}}) & =\mathbb{E}[\big(y_{i}-Q_{i}(\mathbf{s},a_{i};\boldsymbol{\theta}_{i})\big)^{2}].\\
\end{split}
\vspace{-.5em}
\end{equation}
Before the weight update process takes place, the total loss is calculated as the the mean across all arms
\vspace{-.5em}
\begin{equation}\label{eq:Q_update_total}
\begin{split}
L(\boldsymbol{\theta}) & =\mathbb{E}\bigg[\frac{1}{N}\sum_{i=1}^{N}L(\boldsymbol{\theta_{i}})\bigg].\\
\end{split}
\vspace{-.8em}
\end{equation}

Then, by differentiating $L(\boldsymbol{\theta})$ w.r.t. $\boldsymbol{\theta}$, weights of the DNN are updated for the next iteration, where $\alpha$ is the earning rate. 
\vspace{-.5em}
\begin{equation}
\boldsymbol{\theta} \leftarrow \boldsymbol{\theta} - \alpha\nabla_{\boldsymbol{\theta}}L(\boldsymbol{\theta}).
\vspace{-.5em}
\end{equation}

Since the gradient descent updates of $\boldsymbol{\theta}$ is different from canonical supervised learnings because the training target $y_{i} = R(\mathbf{s},\mathbf{a})+\gamma\max_{a^{\prime}_{i} \in \mathbf{\Omega}_{i}}Q_{i}(\mathbf{s}^{\prime},a^{\prime}_{i};\boldsymbol{\theta}_{i})$ is generated by the same $Q$-function $Q_{i}(s,a_{i};\boldsymbol{\theta}_{i})$ that is being trained.  Therefore, to improve stability and performance of the training process, we improve the training algorithm in the following three ways. 

\emph{\textbf{1)}}. We maintain a delayed version of the $Q$-function, $Q_{i}(\mathbf{s}^{\prime},a^{\prime}_{i};\boldsymbol{\theta}^{\prime}_{i})$, for the estimation of the maximum next state reward, which was proposed~\cite{mnih2015human} to improve the stability of their DQN. As such, the target function in (\ref{eq:target}) is updated to 
\vspace{-.6em}
\begin{equation}\label{eq:target_updated}
y_{i} = R(\mathbf{s},\mathbf{a})+\gamma\max_{a^{\prime}_{i} \in \mathbf{\Omega}_{i}}Q_{i}(\mathbf{s}^{\prime},a^{\prime}_{i};\boldsymbol{\theta}^{\prime}_{i}).
\vspace{-.6em}
\end{equation}
The delayed $Q$-function is updated with the newest weights every $C$ (``target sync gaps") steps by setting $\boldsymbol{\theta}^{\prime}_{i}  \leftarrow \boldsymbol{\theta}_{i}$. 

\emph{\textbf{2)}}. To overcome the overestimation of action values, we implement \emph{Double Q-learning}\cite{hasselt2010double} to address the positive bias in estimation introduced when the maximum expected action values are instead approximated by the maximum action values in $Q$-learning. Specifically, we use the up-to-date $Q$-function $Q_{i}(\mathbf{s},a_{i};\boldsymbol{\theta}_{i})$ to determine the optimal sub-actions, i.e.,
\vspace{-.6em}
\begin{equation}\label{eq:optimal_sub_action}  
a^{\prime*}_{i} = \arg\max_{a^{\prime}_{i} \in \mathbf{\Omega}_{i}}Q_{i}(\mathbf{s}^{\prime},a^{\prime}_{i};\boldsymbol{\theta}_{i}). 
\vspace{-.6em}
\end{equation}
The accumulated reward of the returned action $a^{\prime*}_{i}$ is estimated by the delayed $Q$-function using (\ref{eq:target_updated}). Therefore, the target in (\ref{eq:target}) is further improved to be
\vspace{-.6em}
\begin{equation}\label{eq:target_up_updated}
y_{i} = R(\mathbf{s},\mathbf{a}) + \gamma Q_{i}(\mathbf{s}^{\prime},\arg\max_{a^{\prime}_{i} \in \mathbf{\Omega}_{i}}Q_{i}(\mathbf{s}^{\prime},a^{\prime}_{i};\boldsymbol{\theta}_{i});\boldsymbol{\theta}^{\prime}_{i}).
\vspace{-.6em}
\end{equation}

\emph{\textbf{3)}}. We implement the ``replay memory" \cite{lin1993reinforcement} where the agent's past experiences are stored (\emph{matrix $\mathcal{D}$}) and might be used more than once for training. See Section~\ref{training_algorithm} for details.

\subsection{Details of the DNN}\label{DDN}
The structure of MACS is demonstrated in the example in Fig.~\ref{fig:DRL_example}, which can be categorised as a \emph{Multilayer Perceptron (MLP)} \cite{kumar2011multilayer}. The input to the MLP is of dimension $N$, which corresponds to a state of the formulated MDP (see Section~\ref{sec:MDP_formulation}). The input layer, which precedes and is fully connected to all action arms and the state layer, consists of 2 hidden fully connected layers with 512 and 256 neurons, respectively. Every action arm contains a hidden layer of 128 neurons which is followed by an output layer with 2 neurons that output the advantage estimations for 2 sub-actions, respectively. The state layer has a hidden layer of 128 neurons followed by a single neuron in its output layer, which gives the estimation of state value. Overall, the output of the MLP is a vector of $Q$-values whose dimension is $2N$. Note that the state value (output of the state layer) and the advantages of all sub-actions (outputs of the all action arms) are combined according to (\ref{eq:aggregation_method}). When the trained MLP is used for making action predictions, the chosen sub-action for each action arm is decided by comparing $Q$-values of its two sub-actions, whichever is greater gets picked, as demonstrated in Fig.~\ref{fig:DRL_example}. In cases where the number of arms giving ``1" output 
is larger than the given synchronization budget, those sub-actions with the greater $Q$-values get picked first until the budget is reached.

The MLP is realised using Keras \cite{chollet2018keras} model with TensorFlow\cite{abadi2016tensorflow}, in which Adam is chosen as the optimiser with initial learning rate of $0.0001$ and Rectified Linear Unit (ReLU)\cite{nair2010rectified} is employed as activation functions for all neurons except for the outputs. The discount factor is set to $\gamma = 0.99$; while the target network is updated every $20$ steps (i.e., $C=20$).
\setlength{\textfloatsep}{0pt}
\begin{algorithm}[tb] 
	\small
	\SetKwInOut{Input}{input}\SetKwInOut{Output}{output}
	\Input{MLP model settings; distributed SDN settings; simulation program for generating rewards.}
	\Output{Trained parameterized $Q$-functions for all arms.}
	Initialize $Q$-functions $Q_{i}(\mathbf{s},a_{i};\boldsymbol{\theta}_{i}), i\in\{1,\dots,N\}$ by instantiating the MLP; Set initial state $s_{0}$; $t = 0$;\\
	Initialize matrix $\mathcal{D}$ with past $(\mathbf{s},\mathbf{a},R(\mathbf{s},\mathbf{a})-\rho\iota,\mathbf{s}^{\prime})$ tuples; \\
	Initialize the delayed $Q$-functions $\boldsymbol{\theta}^{\prime}_{i} \leftarrow \boldsymbol{\theta}_{i}, i\in\{1,\dots,N\}$. \\
	\While{ $t\leq T$}
	{		
		\ForEach {time instant $t$}
		{ 	$\vars{indicator} = 0,1$ with probabilities      $1-\epsilon,\epsilon$, respectively; \\
			\eIf{$\vars{indicator} = 0$}
			{
				Select an action $a_{t}$ randomly;
			}
			{
				Select an action $a_{t}$ according to  Section~\ref{DDN};
			}		    
			Pass on the $(\mathbf{s}_{t},\mathbf{a}_{t})$ to the simulation program to get return $r_{t}$ and $\mathbf{s}_{t+1}$ ;\\
			Store $(\mathbf{s}_{t},\mathbf{a}_{t},r_{t}-\rho\iota,\mathbf{s}_{t+1})$ in $\mathcal{D}$;\\
			Pull minibatch $\mathcal{D}_{t}$ of $(\mathbf{s}_{i},\mathbf{a}_{i},r_{i}-\rho\iota,\mathbf{s}_{i+1})$ from $\mathcal{D}$;
			
			\ForEach {$(\mathbf{s}_{i},\mathbf{a}_{i},r_{i}-\rho\iota,\mathbf{s}_{i+1})$ in $\mathcal{D}_{t}$ }
			{   \ForEach {$ g \in\{1,\dots,N\}$}
				{Calculate $L(\boldsymbol{\theta}_{g})$ from $a_{g}$ and $y_{g}$ using (\ref{eq:Q_update}), (\ref{eq:optimal_sub_action}), and (\ref{eq:target_up_updated}), respectively;
				}
				Calculate aggregrated loss $L(\boldsymbol{\theta})$ according to (\ref{eq:Q_update_total});\\
				Update weights: $\boldsymbol{\theta} \leftarrow \boldsymbol{\theta} - \alpha\nabla_{\boldsymbol{\theta}}L(\boldsymbol{\theta})$;\\
			}
			
			\If{$t\mod C = 0$ (\texttt{C: target sync gap})}
			{   \ForEach {$ i \in\{1,\dots,N\}$}
				{
					Update weights: $\boldsymbol{\theta}^{\prime}_{i} \leftarrow  \boldsymbol{\theta}_{i}$.
				}
				
			}
		}
	}
	\caption{Training algorithm for MACS}
	\label{alg:aglorithm1}	
\end{algorithm}
\normalsize

\subsection{The Training Algorithm}
\label{training_algorithm}
So far, the design of immediate reward only takes into account the average reduction in service latency after synchronization broadcastings. Since most broadcastings of up-to-date BIS bring positive rewards, 
this makes the agent think that it is always better to broadcast as many as possible. However, as we have the synchronization budget constraint, we need to make this realisable to the agent during training. To this end, we offset the immediate reward defined in the MDP by a small value for the sub-actions where up-to-date BISes are broadcasted (i.e., sub-actions indicated by $1$, referred to as \emph{positive sub-actions}). In particular, let $\iota$ denote the unit offset for each positive sub-action, and $\rho$ be the number of positive sub-actions for the state-action pair $(\mathbf{s},\mathbf{a})$; then, the 4-tuple stored is  $(\mathbf{s},\mathbf{a},r,\mathbf{s}^{\prime})$, where $r = R(\mathbf{s},\mathbf{a})-\rho\iota$.

MACS is first pre-trained on past $(\mathbf{s},\mathbf{a},r,\mathbf{s}^{\prime})$ tuples already stored in fixed-size matrix $\mathcal{D}$ (i.e., the agent's ``reply memory"). Then, MACS starts making synchronization decisions, while keeps being trained in a semi-online fashion in the sense that new $(\mathbf{s},\mathbf{a},r,\mathbf{s}^{\prime})$ tuples gradually replace old entries in matrix $\mathcal{D}$ on a first-in-first-out basis. At each training iteration where $Q$-learning update takes place, minibatch samples of stored experience tuples are pulled randomly from $\mathcal{D}$ for training.

The training process is summarised in Algorithm~\ref{alg:aglorithm1}. 

\section{Evaluation}
\label{sec:evaluation}
This section starts by introducing the evaluation scenarios and benchmarks in Section~\ref{net_dynamic_model}. Then, network settings for evaluation scenarios are described in Section~\ref{net_settings}. Finally, we present evaluation results and analysis in Section~\ref{evaluation_results}.

\subsection{Evaluation Scenarios and Performance Benchmarks}
\label{net_dynamic_model}

Three scenarios are considered for evaluations, where \emph{Scenario~1} serves as the baseline where the overall performance of MACS is compared against other benchmarks. In addition, under the settings of Scenario~1, we also evaluate the performance of MACS in a fully online manner, for which no pre-training on history data is performed. Then, \emph{Scenario~2} and \emph{Scenario~3} evaluate the sensitivity of MACS performance under different network settings.  Specifically, we compare the performance of MACS alongside other benchmarks under varying  BIS value distributions in Scenario~2 and varying synchronization budget in Scenario~3. In the following, we briefly discuss the benchmarks used for comparing the performance of MACS. 


\emph{\textbf{The Full/Worst Controller Synchronization Levels}} correspond to the case where all up-to-date BISes are broadcast to domain controllers at every time slot; and the case where there is no broadcasting of any up-to-date BIS at any time slot, respectively. We can see that the full synchronization level is identical to having one logical central controller and that it incurs the maximum synchronization overheads. 
Note that these two cases serve as the lower and upper bounds of our performance metric and they are not subject to the given synchronization budgets. 

\emph{\textbf{The Greedy (MinMax) Algorithm}} is a simple controller synchronization scheme that aims at reducing the staleness of controller-perceived BIS values by minimising the maximum state value of the defined MDP. Specifically, with the given synchronization budget at a time slot, the up-to-date values of those BISes that have not been synchronized to all domain controllers the longest get broadcasted first. 

\emph{\textbf{The Anti-entropy \cite{muqaddas2016inter} Algorithm,}} which is implemented in the ONOS controller\cite{berde2014onos}, is based on a simple gossip algorithm that controllers randomly synchronize with each other\cite{muqaddas2016inter}. W.r.t. the definition of controller synchronization in our problem (Definition~\ref{def:synch}) and the given synchronization budget, controller synchronization with the anti-entropy algorithm is carried out such that up-to-date values of BIS are randomly selected for synchronization broadcastings.
\begin{table}
	\centering
	\caption{Evaluation Parameters}
	\begin{tabular}{M{0.32\linewidth}| M{0.15\linewidth} |M{0.19\linewidth} |M{0.14\linewidth}}
		\hline
		Parameter & Scenario $1$ & Scenario $2$ & Scenario $3$
		\\ \hline
		Synchronization budget distribution  & Poisson distributed with $\lambda = 3$ & Poisson distributed with $\lambda = 3$& Poisson distributed with $\lambda = 1,3,5$
		\\ \hline
		BIS value distribution  & Uniformly distributed & Gaussian distributed with $\mu = 10$, $\sigma = 5,8,11$ & Uniformly distributed 
		\\ \hline
		Probabilities that BISes change value & \multicolumn{3}{c}{\makecell{The probability of BIS $i$ changing value \\is proportional to  $\frac{1}{{\sigma \sqrt {2\pi } }}e^{{{ - \left( {i - \mu } \right)^2 } \mathord{\left/ {\vphantom {{ - \left( {x - \mu } \right)^2 } {2\sigma ^2 }}} \right. \kern-\nulldelimiterspace} {2\sigma ^2 }}},$  \\  $(\mu = 30, \sigma = 10)$}}
		\\ \hline
		Probabilities that service installed in domains & \multicolumn{3}{c}{ \makecell{$30\%$ probability in any $4$ domains and $70\%$ \\probability in the other $4$ domains}}
		\\ \hline 
		Service request pattern  & \multicolumn{3}{c}{\makecell{The probability that service $i$ is requested is \\proportional to $(q+i)^{-\beta} (q =5, \beta = 0.8)$}}
		\\ \hline
		The number of domains & \multicolumn{3}{c}{$8$ domains} 
		\\ \hline
		The number of services & \multicolumn{3}{c}{\makecell{$10$ unique services, each installed \\twice in $2$ different domains}} 
		\\ \hline
	\end{tabular}\label{tb:parameters}
\end{table}

\begin{figure*}
	\smallskip
	\centering
	\begin{subfigure}[b]{0.32\textwidth}
		\centering
		\includegraphics[width=\linewidth,height=0.6\linewidth]{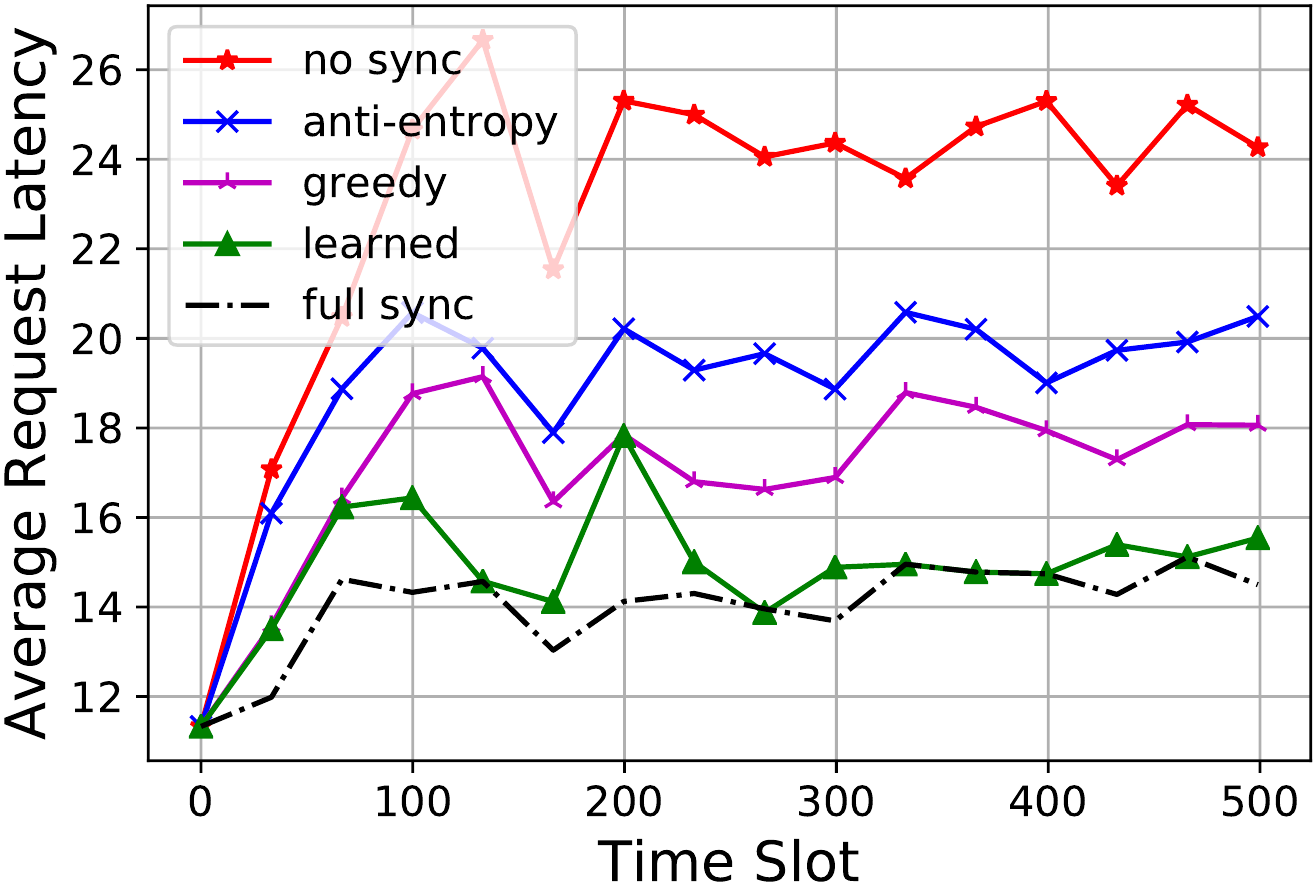}
		\caption{Scenario~1: Average request latency of all constructed paths at all time slots.}
		\label{fig:s2_1}
	\end{subfigure}
	\begin{subfigure}[b]{0.32\textwidth}
		\centering
		\includegraphics[width=\linewidth,height=0.6\linewidth]{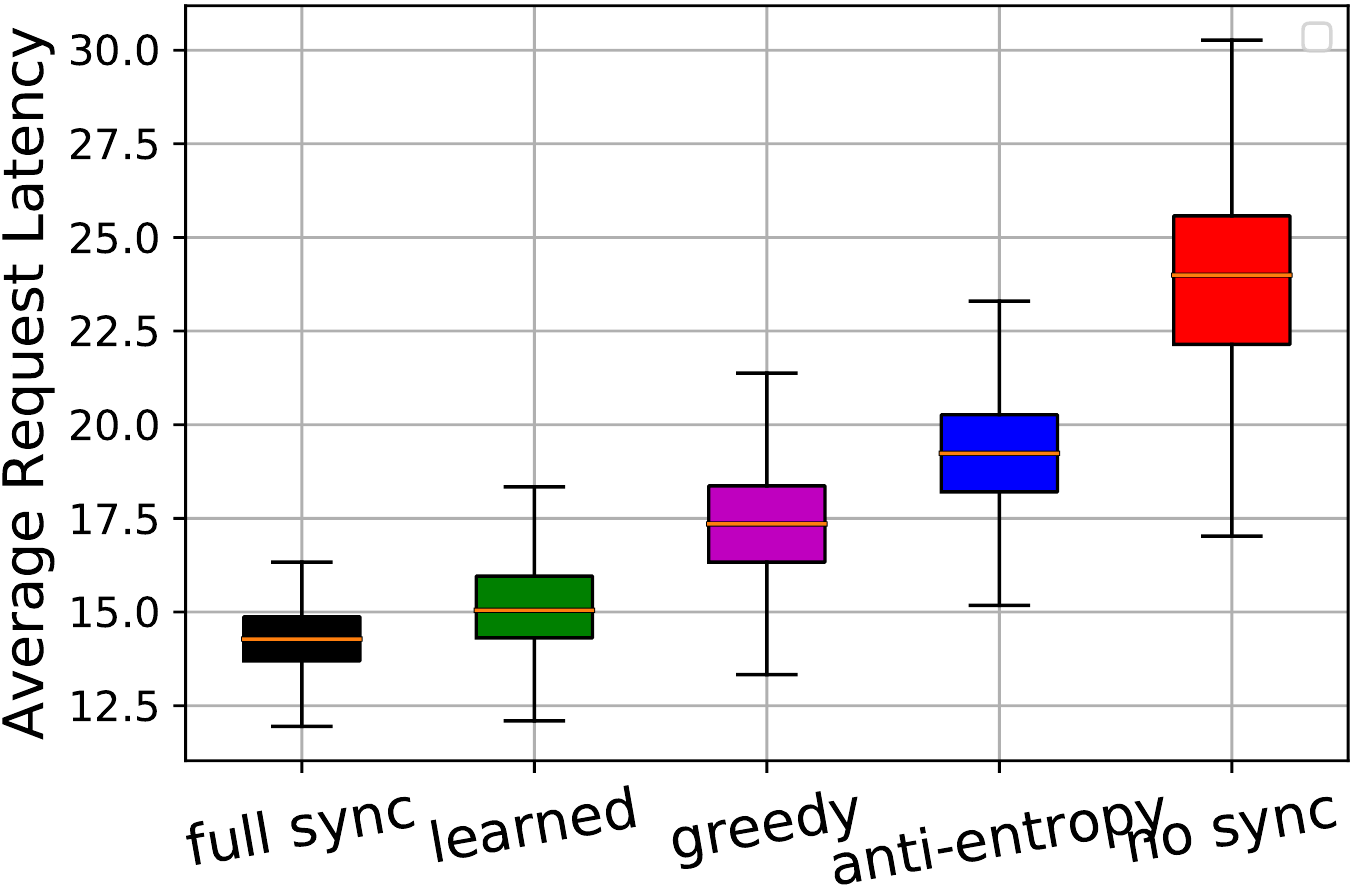}
		\caption{Scenario~1: Box plot of average request latency over all times slots.}
		\label{fig:s2_3}
	\end{subfigure}   
	\begin{subfigure}[b]{0.32\textwidth}
		\centering
		\includegraphics[width=\linewidth,height=0.63\linewidth]{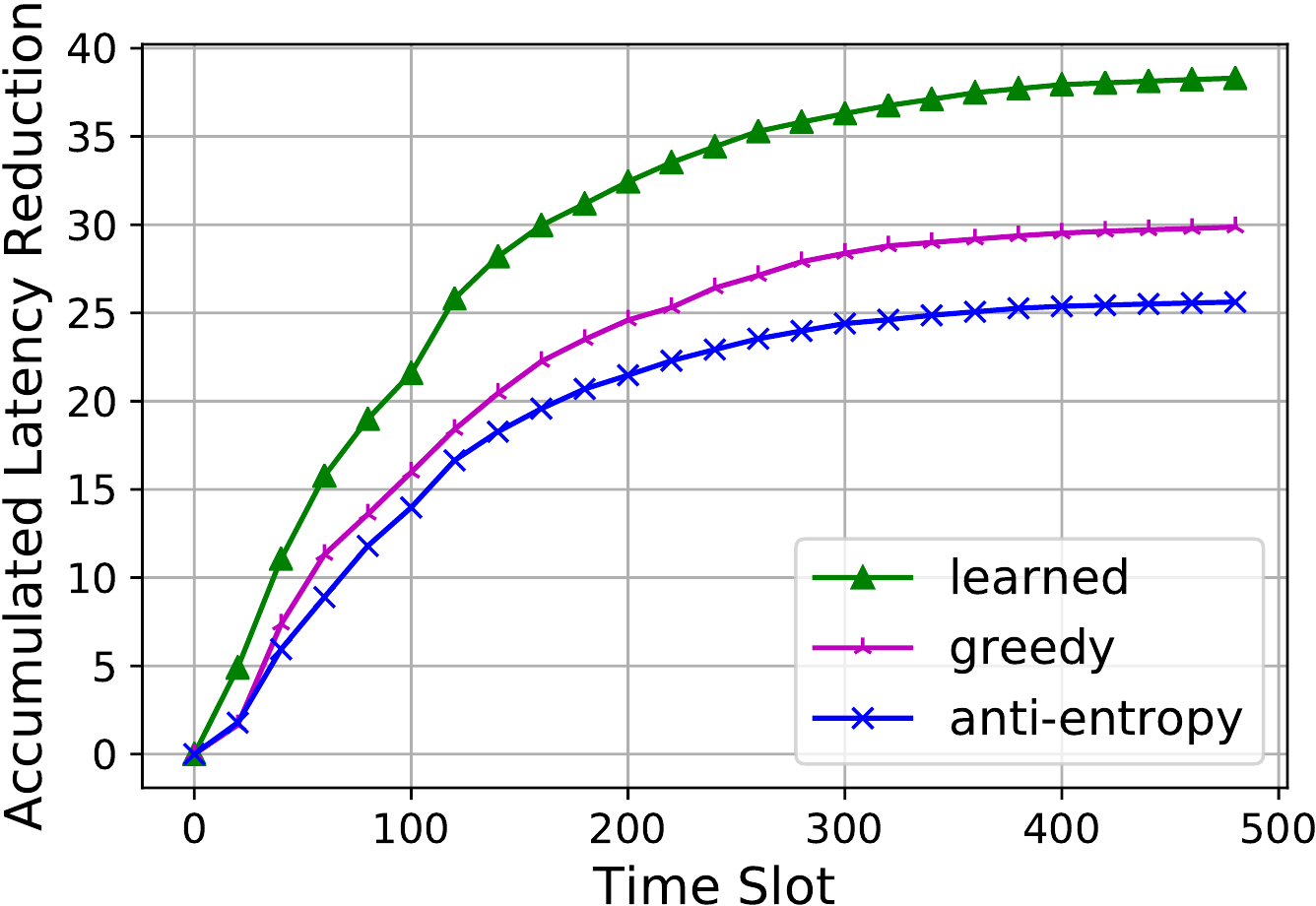}
		\caption{Scenario~1: Accumulated average request latency reductions over time slots.}
		\label{fig:s2_4}
	\end{subfigure}
%
	\begin{subfigure}[b]{0.32\textwidth}
		\centering
		\includegraphics[width=\linewidth,height=0.6\linewidth]{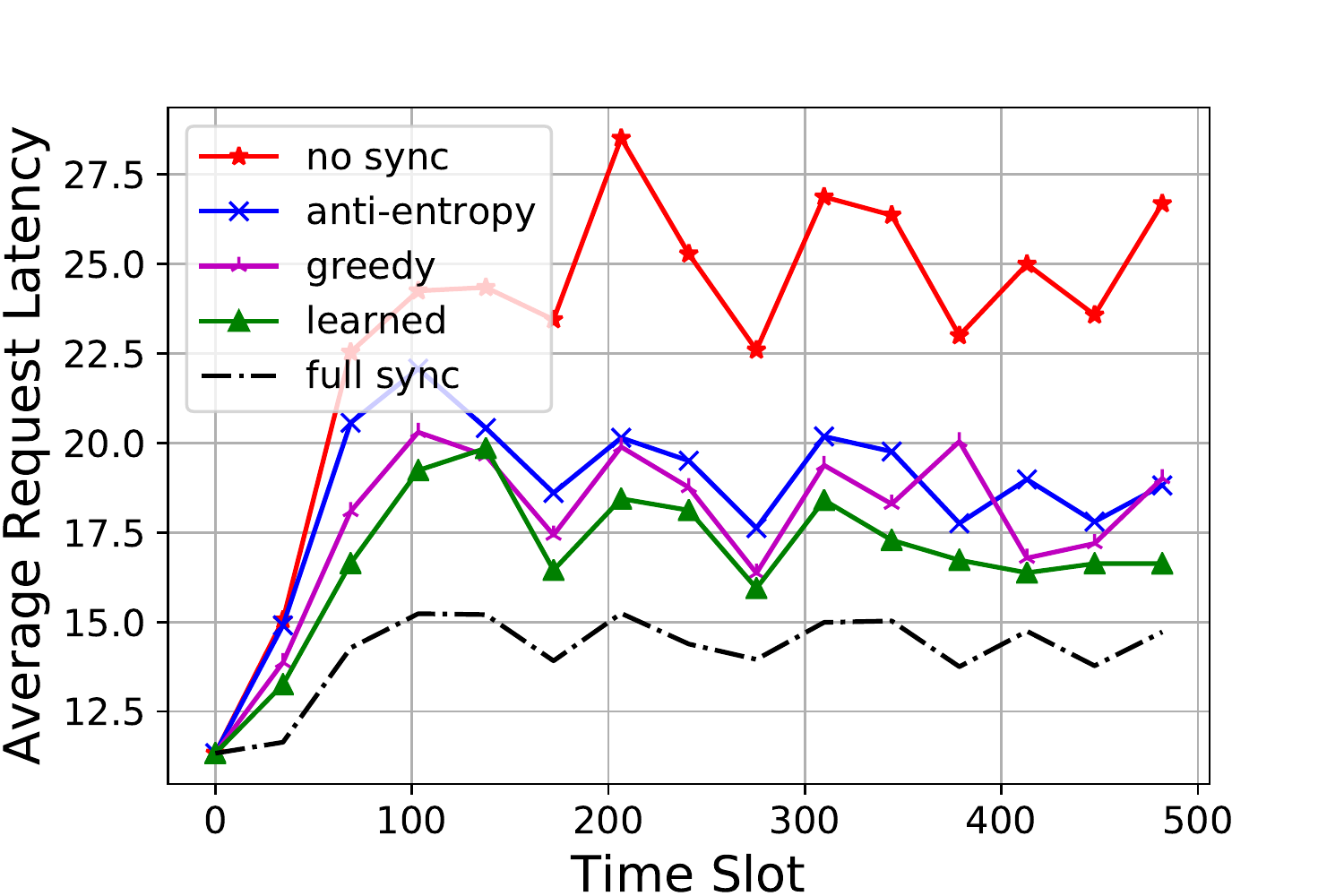}
		\caption{Scenario 1: Average request latency at all time slots with \emph{online} MACS.}
		\label{fig:s1_1}
	\end{subfigure}
	\begin{subfigure}[b]{0.32\textwidth}
		\centering
		\includegraphics[width=\linewidth,height=0.6\linewidth]{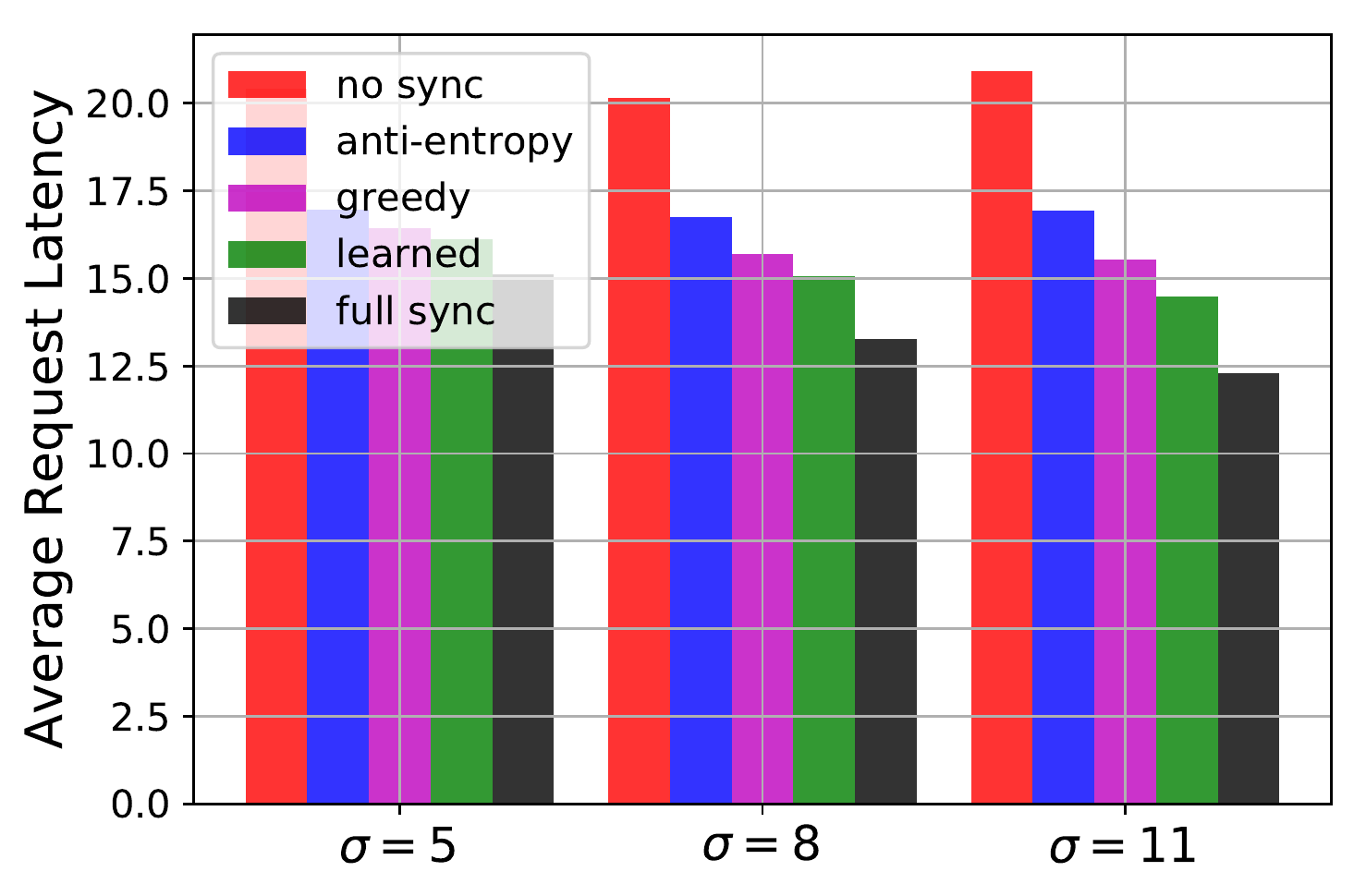}
		\caption{Scenario 2: Average request latency with different BIS value distributions.}
		\label{fig:s1_3}
	\end{subfigure}   
	\begin{subfigure}[b]{0.32\textwidth}
		\centering
		\includegraphics[width=\linewidth,height=0.6\linewidth]{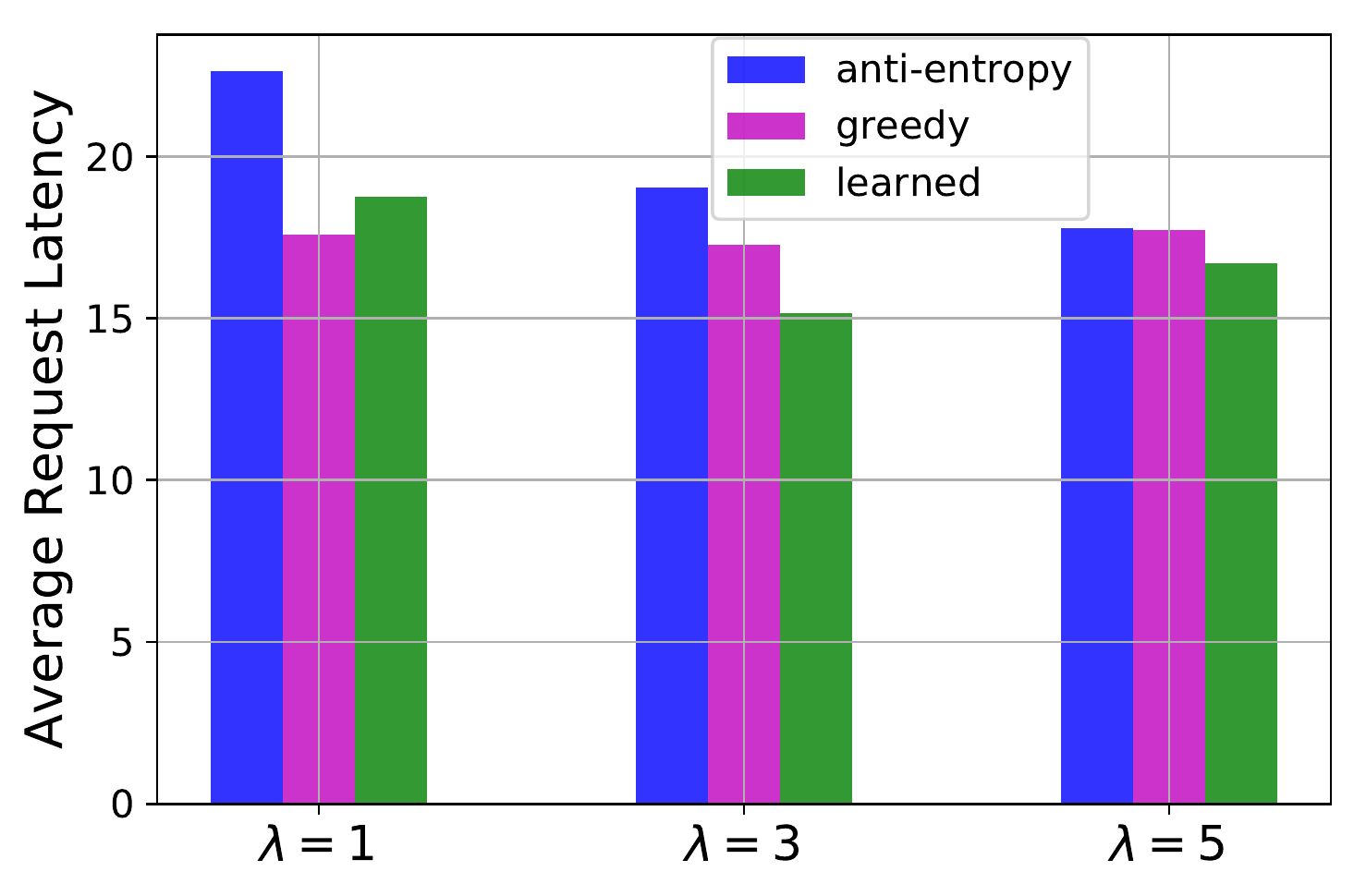}
		\caption{Scenario 3: Average request latency with different synchronization budget distributions.}
		\label{fig:s1_4}
	\end{subfigure}
	\caption{Evaluation results.}
	\label{fig:results_scenario1}   
	\vspace{-2em} 
\end{figure*}

\subsection{Network Settings}
\label{net_settings}
\subsubsection{Network topology of the simulated SDN network}
For our evaluations, all domain-wise topologies, i.e., how domains are connected to each other, are constructed according to a real dataset. Specifically, there are $8$ domains connected according to the \emph{degree distribution} extracted from the ``CAIDA AS-27524" dataset \cite{Snap}, which refers to the distribution of the number of neighbouring domains of an arbitrary domain. 
\subsubsection{Server distribution in domains and user request pattern} $10$ unique services are considered in all scenarios, with two installations in two different domains for each service. Moreover, domains are divided randomly and equally into two groups, service installations are more likely to be inside the first group (with $70\%$ probability) than the second one (with $30\%$ probability). In addition, service request patterns follow Zipf-Mandelbrot distribution\cite{borst2010distributed}, which is widely used to model the content popularity in content delivery networks (CDN). Specifically, the popularity of $i$-th most popular service is proportional to $(q+i)^{-\beta}$, where $q=5$ and $\beta = 0.8$. 
\subsubsection{Network dynamicity pattern} The distribution of available synchronization budget for different time slots is modelled by Poisson process with mean $\lambda = 3$ in Scenario~1 and Scenario~2; whereas in Scenario~3, three experiments are conducted with $\lambda = 1$, $\lambda = 3$, and $\lambda = 5$ to evaluate the impacts of the level of available budget. As for BIS values, they are uniformed distributed in Scenario ~1 and Scenario~3.\footnote{In particular, BISes are randomly drawn from the set $\{1,2,4,6,8,13,17,20,25,30\}$. Note that there is no assumption on the relationship between new and old BIS values (e.g., whether or not they are correlated).} In order to evaluate the impact of BIS value distribution on the performance of MACS, we conduct three experiments where the BIS values are Gaussian distributed with mean $\mu=10$ and standard deviation (STD) being $\sigma = 5$, $\sigma = 8$, and $\sigma = 11$, respectively. All evaluation settings are summarised in Table~\ref{tb:parameters}.

\emph{\textbf{Remark:}} It should be noted that the network settings described are only for the sake of evaluations, there are no assumptions on any of the parameters used above.

\subsection{Evaluation Results}
\label{evaluation_results}

All evaluation results are presented in Fig.\ref{fig:results_scenario1}. In particular, Fig.\ref{fig:s2_1} shows plots of average request latencies of all constructed service paths in each time slot. Fig.\ref{fig:s2_3} shows the box plots of average request latencies resulted from different synchronization algorithms. Fig.\ref{fig:s2_4} contains plots of accumulated latency reduction defined in (\ref{eq:long_term_reward}), which is the maximisation objective of the MDP. Fig.\ref{fig:s1_1} is similar to Fig.\ref{fig:s2_1}, except that here MACS is trained in the fully online manner as synchronization decisions are made and new $(\mathbf{s},\mathbf{a},r,\mathbf{s}^{\prime})$  tuples become available for training.  Fig.\ref{fig:s2_1}-Fig.\ref{fig:s2_4} all correspond to Scenario~1. The bar plot in Fig.\ref{fig:s1_3} are the evaluation results of Scenario~2 where we vary the STD of BIS value distributions. Fig.\ref{fig:s1_4} presents the evaluation results of Scenario~3, which illustrate the impacts of varying available synchronization budget distributions. Note that the bar plots in Fig.\ref{fig:s1_3}-Fig.\ref{fig:s1_4} are the averaged results over all time slots. In these figures, the legend ``learned'' refers to the synchronization policy learned via MACS; ``full sync",  ``greedy", ``anti-entropy", and ``no sync" respectively refer to the  synchronization cases discussed in Section~\ref{net_dynamic_model}. Moreover, it should be stressed that in these plots, the unit of the performance metric, i.e., the request latency recorded in the simulated network, and the time scale of the MDP, i.e., the time slot, are different as discussed in Section~\ref{sec:MDP_formulation}.


\emph{1) Superiority of MACS:} From Fig.\ref{fig:s2_1}, we can see that the gap between ``full sync" and ``no sync" curves clearly demonstrates the important role controller synchronization plays in improving performance. Among the three synchronization algorithms implemented, MACS consistently performs the best in both scenarios. Surprisingly, after $300$ time slots, MACS, which runs on a limited synchronization budget, can almost achieve full sync performance, as can be seen in Fig.\ref{fig:s2_1} where two curves overlap. The superiority of MACS over the other two schemes is also evident in Fig.\ref{fig:s2_3}, which compares the statistic properties of the request latency results of different synchronization regimes. In addition, these results reveal that while the goal is to minimise long-term average request latency, this objective is achieved by consistently minimising average request latency at each time slot. 

\emph{2) Superiority of MACS for maximising the Bellman equation:} Recall that our objective is to maximise the accumulated reductions in request latency over a period of time, i.e., to maximise $V^{\pi}(\mathbf{s}_0) = \mathbb{E}_{\pi}\bigg[\sum_{t=1}^{T}\gamma^t R(\mathbf{s}_{t},\mathbf{a}_{t})|\mathbf{s}_0\bigg]$. The evaluation results in  and Fig.\ref{fig:s2_4} confirm the superiority of MACS in achieving this goal. In particular, during the testing period of $500$ time slots, MACS outperforms the greedy algorithm by approximately $30\%$; the anti-entropy by $56\%$, respectively.

\emph{3) Online performance of MACS:} In Fig.\ref{fig:s1_1}, it can be seen that although the synchronization policy by MACS outperforms that of other algorithms, the performance margin is significantly smaller than in Fig.\ref{fig:s2_1} where pre-training is conducted. The performance differences here indicate the value of pre-training with history data before using MACS to generate synchronization policies.  From Fig.\ref{fig:s1_1} we can also see that as the training continues and more training samples become available, synchronization decisions by MACS keep improving.

\emph{4) The impact of BIS value distribution:} From the results in Fig.\ref{fig:s1_3} we can see that when the STDs of the BIS value distributions increase from 5 to 11, only  the ``full sync'' and ``learned'' latency results delivered by MACS show improvements. This suggests that the policies developed by MACS are more valuable when the network conditions are highly volatile, manifested by a wider range where BIS values can vary.  In comparison,  the anti-entropy and greedy algorithms are non-adaptive to varying network conditions, as expected.

\emph{5) The impact of available synchronization budget:} Since the ``full sync'' and ``no sync'' cases are not subject to the synchronization budget constraints, we do not plot their evaluation results in  Fig.\ref{fig:s1_4}. We can see that when the average synchronization budget ($\lambda$ in the Poisson process) at a time slot increases from 1 to 5, the performance of MACS initially improves and then deteriorates. Recall that MACS makes synchronization decisions by selecting the sub-actions with the highest $Q$-values first until the synchronization budget is exhausted. Therefore, when a large synchronization budget is allowed, MACS may select some actions with low or even negative $Q$-values, which explains the performance deterioration. On the other hand, when the synchronization budget is too constraints (i.e., $\lambda = 1$), the greedy algorithm outperforms MACS. This shows the important role of the budget in regulating the controller synchronization process. Moreover, the greedy algorithm appears to be insensitive to budget levels, i.e., having greater synchronization budget does not help its performance.



\section{Related Work}
\label{sec:related_work}
\subsubsection{Distributed SDN}
Among a couple of  design proposals of distributed SDN  controller architecture,  OpenDaylight~\cite{opendaylight} and ONOS~\cite{berde2014onos} are two state-of-the-art SDN controllers proposed to realise logically centralised but physically distributed SDN architecture. In addition, controllers such as Devoflow~\cite{curtis2011devoflow} and Kandoo~\cite{hassas2012kandoo} are designed with their specific aims. However, most of these controller architectures do not emphasize or justify detailed controller synchronization protocols they employ. 

\subsubsection{Controller Synchronizations}
Most existing works on controller synchronization assume either strong or eventual consistency models\cite{sakic2018impact}. The authors in \cite{levin2012logically} show that certain network applications can rely on the eventual consistency to deliver acceptable performance. This work \cite{guo2014improving} shows how to avoid network anomalies such as forwarding loops and black holes under the eventual consistency assumption. Similar to our approach, the works \cite{aslan2016adaptive,poularakis2019learning,sakic2017towards} propose dynamic adaptation of synchronization rate among controllers. Compared to these works, MACS is much more generic and versatile in that there are no assumptions or constraints on any network parameters due to its model-free nature.   

\subsubsection{Reinforcement Learning in SDN}
Some recent high-profile successes~\cite{mnih2015human,silver2016mastering} attract enormous interests in using RL techniques to solve complicated decision-making problems. In the context of SDN, the authors in \cite{zhang2018q} apply RL-based algorithms to solve service placement problem on SDN switches. A routing-focused controller synchronization scheme is developed using DRL-based approaches in \cite{DBLP:journals/corr/abs-1812-00852}, where the MDP the authors formulated is easier to solve, as they assume an uniform synchronization budget and coarse-grained synchronization decisions where the decisions are at the level of controller pairs. This work\cite{lin2016qos} also discusses the routing problem in SDN using RL techniques. However, the discussion is only limited to intra-domain routing under strong assumptions on the network topology. In addition, tabular settings are used in this work without generalisations. 
\section{Conclusion}
\label{sec:conclusion}
In this paper, we investigated the controller synchronization problem with limited synchronization budget in distributed SDN, for which our aim was to find the policy that maximises the performance enhancements brought by controller synchronizations over a period of time. We formulated the controller synchronization problem as an MDP which has a large state-action space. We identified challenges in solving the formulated MDP and designed a DRL-based algorithm, called MACS, which absorbs various DNN design principles to tackle these challenges. Due to the DNN's ability in learning the network dynamicity patterns which results in near optimal use of the given limited synchronization budget, evaluation results showed that MACS consistently outperforms state-of-the-art SDN controller synchronization algorithms/heuristics operating under the same synchronization budget by large margins.  

\bibliographystyle{IEEEtran}
\bibliography{reference}

\vspace{12pt}

\end{document}